# Competitive Solvation of the Imidazolium Cation by Water and Methanol

Vitaly Chaban[1]

1) Instituto de Ciência e Tecnologia, Universidade Federal de São Paulo, 12231-280, São José dos Campos, SP, Brazil

2) Department of Chemistry, University of Southern California, Los Angeles, CA 90089, United States

**Abstract**. Imidazolium-based ionic liquids are widely used in conjunction with molecular liquids for various applications. Solvation, miscibility and similar properties are of fundamental importance for successful implementation of theoretical schemes. This work reports competitive solvation of the 1,3-dimethylimidazolium cation by water and methanol. Employing molecular dynamics simulations powered by semiempirical Hamiltonian (electronic structure level of description), the local structure nearly imidazolium cation is described in terms of radial distribution functions. Although water and methanol are chemically similar, water appears systematically more successful in solvating the 1,3-dimethylimidazolium cation. This result fosters construction of future applications of the ternary ion-molecular systems.

**Key words**: ionic liquid, imidazolium, water, methanol, solvation, hydration, semiempirical, PM7-MD

---

[1] E-mail: vvchaban@gmail.com

**Research Highlights**

1. Solvation and hydration of the imidazolium cation are reported.

2. The derived radial distribution functions are helpful to interpret experimental data.

3. Molecular trajectories are propagated using simplified electronic structure description.

TOC Image

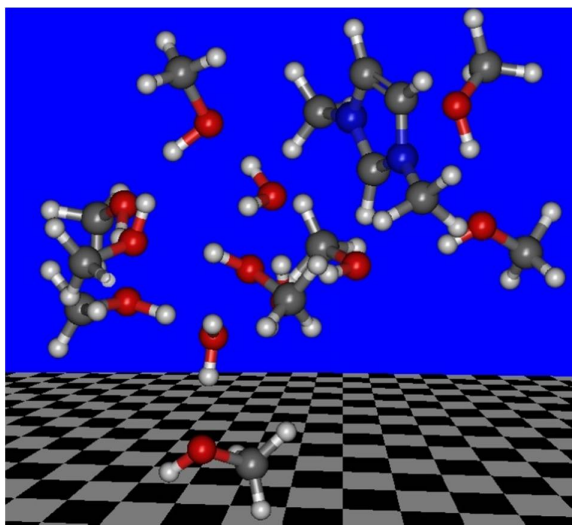

**Introduction**

Room-temperature ionic liquids (RTILs)[1] constitute an actively researched area nowadays.[2-19] The interest to RTILs is evidenced by a wide range of their existing and prospective applications in laboratory chemistry and chemical technology. Low vapor pressure and toxicity (of certain species), high thermal stability and electrical conductivity, and – most importantly – tunability of structure are the properties, which justify an existing interest to this field in the chemical, physical, and biological communities.[1,17,19-24] Imidazolium-based RTILs have been probably the best studied family thus far.[4-19,25-37] Both direct physical investigations[2,3,9-11,13,17,18,30] and various sorts of numerical computer simulation[1,16,22,23,38,39] have been actively applied to these ionic liquids.

The imidazolium-based RTILs are excellently soluble in most polar solvents[1,2,11,13,15,17,20-32,38-44] and are relatively cheap to synthesize. Polar molecular co-solvents, such as water, methanol, acetonitrile, acetone, n-butanol, alter physical and chemical properties of the imidazolium-based RTILs drastically.[35,45-47] This occurs thanks to energetically favorable interaction of polar moiety with the imidazole ring. In particular, hydrogen bonding is frequently observed between intrinsically acidic hydrogen atom, H5, of the imidazole ring and the most electron-rich atom of the co-solvent. Usually, this atom is either oxygen or nitrogen.[45,46]

To design any process involving RTIL on an industrial scale, a range of physical properties must be known, including shear viscosity, density, surface tension, melting and boiling temperatures. Suitable set of properties will enhance liquid-liquid extraction, gas absorption, distillation, condensation, phase separation and so on. The corresponding data in the case of imidazolium-based RTILs are abundant; however, they are sometimes contradictory.[42] A series of recent works have been devoted to reporting various properties of binary and ternary systems involving imidazolium-based ionic liquids, water, methanol, acetonitrile, and certain additional polar molecular co-solvents.[26,30,40,41,45,48] It is now clear that molecular co-solvents immobilize bulky organic ions, decrease share viscosity of pure RTILs, alter the composition of ionic clusters in solution, etc.

Jan et al. reported detailed conductometric studies of 1-butyl-3-methylimidazolium chloride and 1-butyl-3-methylimidazolium hexafluorophosphate in acenonitrile + methanol mixed solvent systems. The study elucidates specific transport properties of the involved ions and solvents, characteristics of ionic association in RTILs, variation of physical properties over a wide range. Such reports are necessary for an optimal usage of the RTIL containing mixed organic solvent systems, which must be employed in high-energy batteries and other electrochemical systems. Furthermore, these results are helpful in better understanding the ion pair effect and organic reaction mechanisms in such organic ion-molecular systems. The temperature and concentration dependences in the case of RTILs are different from the case of small inorganic cations and anions due to a significant role of non-electrostatic interactions and entropy contribution.

Regardless of significant volume of diverse numerical information on the discussed systems, molecular- and atomistic-precision interpretation is scarce. This is a global problem originating from versatile interactions in these complicated systems and decisive role of every constituent, both ionic and molecular ones. Another problem is inability of traditional experimental tools to provide an enough resolution and derive/isolate impacts of each component. In turn, computer simulation techniques can investigate both ion-molecular structures at the ultimate resolution and macroscopic properties (applying relationships of statistical mechanics). However, in practice simulated systems suffer from insufficient complexity. Most computational reports are devoted to one-component systems.[34] An order of magnitude less papers are devoted to two-component systems (ionic liquid plus molecular co-solvent).[33,35,45] Three-component systems[46,48] are still exotic in simulations. Simulations of many-component systems require appropriately parameterized interaction models (known as empirical force fields). Each additional species added to the system increases a cost of force field derivation by a few times, since every pair interaction must be considered separately. The models of different molecules derived independently are sometimes mixed within the same system without re-parameterization and extensive validation. This practice is dangerous, especially when electronic polarization effects play a noticeable role in the considered simulation setup. Electronic polarization is the case in almost every ionic liquid. Obviously, a gap between theory and experiment persists and unfavorably impacts an overall progress of the field.

In the present Letter, I will demonstrate the performance of an alternative computer simulation approach, PM7-MD.[48] The PM7-MD method will be applied to address a problem of competitive solvation of the 1,3-dimethylimidazolium (MMIM$^+$) cation by water and methanol (Figure 1). That is, the three-component system will be investigated. The PM7-MD method[48] does not require parameterization for new molecules and ions as long as the involved chemical elements were parameterized to successfully reproduce electronic structures. The method follows the Born-Oppenheimer approximation, whereby electrons and nuclei move independently. Any energy exchange is forbidden between these species in the course of the simulation. Unlike other simulation methods based on the Born-Oppenheimer approximation, PM7-MD is extremely fast.[48] The trajectories spanning hundreds of picoseconds are accessible for a system containing up to 1 000 electrons. Compare, simulations employing density functional theory (pure exchange-correlation functionals) rarely extend beyond 10 ps and contain at most 100 second-period atoms and hydrogen atoms. It is, nevertheless, important to remember that performance of any electronic structure method heavily depends on how fast the self-consistent field procedure converges. This feature depends on chemical identity of the system and the size of the chosen basis set.

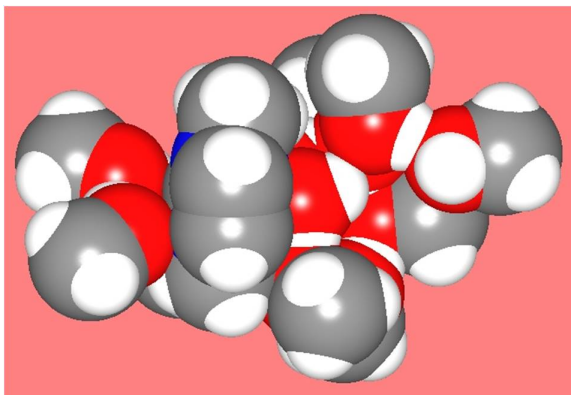

Figure 1. A sample simulated system. The imidazolium cation (q = +1 e) is simultaneously solvated by water and methanol molecules. Oxygen atoms are red, carbon atoms are grey, nitrogen atoms are blue, and hydrogen atoms are white. The system was photographed at the point of its thermodynamic equilibrium. See Table 1 for summarized simulation details.

**Methodology**

Four systems are discussed in this work (Table 1). The first three systems contain the 1,3-dimethylimidazolium cation and certain number of water and methanol molecules representing a mixed polar solvent. Three different concentrations are considered to understand whether the concentration can change solvation regularities. The fourth system contains lithium ion ($Li^+$) instead of $MMIM^+$. This system is provided for comparison of the first solvation shell structure in the case of RTILs and small inorganic ions. All simulations were performed at 333 K to somewhat accelerate molecular dynamics (MD), as compared to room conditions. Note that chosen temperature is lower than the experimentally determined normal boiling point of methanol, 337.85 K.

Table 1. Simulated systems and their fundamental properties. System 4 contains lithium cation instead of imidazolium cation for the sake of comparison. Proper equilibration of all systems was thoroughly controlled by analyzing evolution of many thermodynamic quantities, such as energy components, dipole moments, selected interatomic distances. Note that equilibration of the non-periodic systems occurs significantly faster due to absence of the long-order structure

| # | # $H_2O$ | # MeOH | # atoms | # electrons | Equilibration time, ps | Sampling time, ps |
|---|---|---|---|---|---|---|
| I | 9 | 3 | 79 | 188 | 12 | 90 |
| II | 6 | 6 | 70 | 170 | 13 | 90 |
| III | 3 | 9 | 61 | 122 | 15 | 80 |
| IV | 0 | 5 | 31 | 70 | 9 | 50 |

The PM7-MD method[48] obtains forces acting on every atomic nucleus from the electronic structure computation using the PM7 semiempirical Hamiltonian.[49-52] PM7 is a parameterized Hartree-Fock method, where certain integrals are pre-determined based on the well-known experimental data. Such a solution also allows for effective incorporation of the electron-correlation effects, while preserving a quantum-chemical nature of the method. Therefore, PM7 is able to capture any specific chemical interaction.[52] PM7 is more mathematically fundamental than any existing force field based technique. The derived forces are coupled with initial positions of atoms and randomly generated velocities (Maxwell-Boltzmann distribution). Subsequently, Newtonian equations-of-motions can be constructed and numerically integrated employing one of the available algorithms. This work relies on the velocity Verlet integration algorithm. This integrator provides a decent numerical stability, time-reversibility, and preservation of the symplectic form on phase space. Due to rounding errors and other numerical inaccuracies, total energy of the system is not perfectly conserved, as in any other MD simulation method. Temperature may need to be adjusted periodically by rescaling atomic velocity aiming to obtain the required value. This work employs a weak temperature coupling scheme proposed by Berendsen[53] with a relaxation time of 100 fs, whereas the integration time-step equals to 1.0 fs.

More details of the PM7-MD implementation applied in the present study are described elsewhere.[48] Local structure of the liquid-matter systems can be completely characterized using a set of radial distribution functions (RDFs). Based on the MD trajectories, as described in Table 1, the RDFs were calculated and analyzed using simple in-home tools. Molecular snapshots were conveniently visualized using Gabedit 2.4.[54]

## Results and Discussion

Figure 2-5 provide the most interesting pair RDFs in systems I-IV (Table 1). The well-pronounced peaks at small distances normally indicate a strong solvation and, possibly, a specific intermolecular interaction like hydrogen bonding. These functions contain all necessary information to make conclusions regarding preferential solvation of MMIM$^+$ by these two solvents. The PM7-MD simulations are in concordance with previous insights indicating that the H5 hydrogen atom plays a major role in solvation of MMIM$^+$. The sharp peak at 2.1 Å suggests a moderately strong hydrogen bond between MMIM$^+$ and water forming its first hydration shell. Favorable interaction of water with the methyl group of MMIM$^+$ is also clearly indicated by the corresponding RDF. However, this peak is located only at 3.5 Å. Note that methyl groups connected directly to the electron-deficient imidazole ring pull a significant amount of positive charge density onto themselves. For instance, the Coulson charges on the methyl groups in the present simulation are +0.30e each (the value was averaged based on the population analysis throughout the MD trajectories). If one or both methyl groups are substituted by larger hydrocarbon radicals, positive charge will delocalize making the methyl group less attractive for electrostatic interactions with counter-ions and polar solvents. Despite water and methanol are chemically similar, the corresponding RDFs appear significantly different. Although the first peak is present at 2.1 Å on the H5-OH RDF, it appears much smaller than the second peak at 4.6 Å. One can conclude that most methanol molecules are pushed to the second solvation shell, whereas the first solvation shell is populated by water. Below, we will try to correlate this observation with hydrogen bonding peculiarities being exhibited by water molecules in the vicinity of MMIM$^+$.

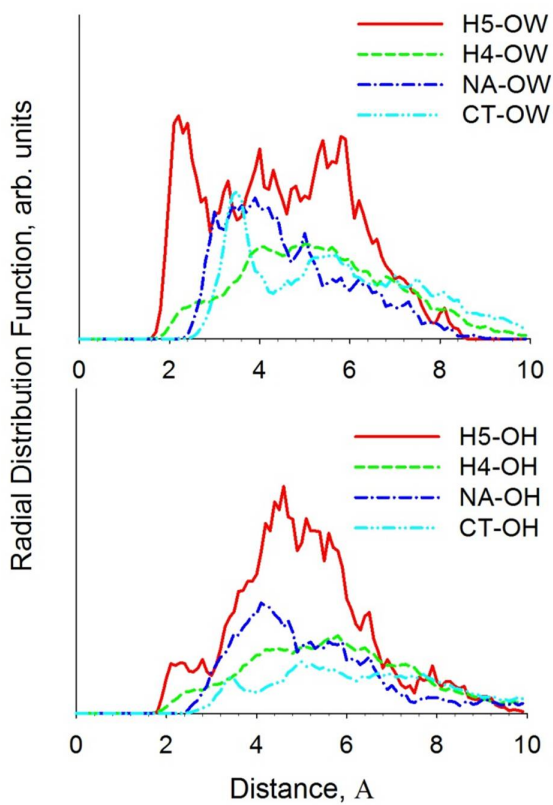

Figure 2. Radial distribution functions computed in system II between various interaction sites of MMIM[+] and the electron-richest atoms (oxygen in both cases) of the solvent molecules. See legends for RDF designation. The atoms are named as in popular empirical force fields: OW – oxygen atom of water; HW – hydrogen atom of water; H5 – intrinsically acidic hydrogen atom of MMIM+ (bonded to a carbon atom between two nitrogen atoms); H4 – two other hydrogen atoms of the imidazole ring; NA – nitrogen atom; CT – carbon atom of the methyl group.

Water occupies the first solvation shell of the imidazolium cation in the equimolar mixture with methanol. Does this trend work at other mixture compositions (Figure3)? The methanol molecules are present in the first solvation shell of MMIM[+] in each system during certain time in the course of MD simulation, but only system III exhibits a little higher first peak for methanol than for water. System III contains 9 methanol molecules and 3 water molecules. Thus, methanol molecules possess more chances to arrange themselves around the cation and push water to the second coordination sphere. All the observed peaks in Figure 3 are relatively diffuse, although H5-O constitutes a primary ion-molecular coordination center (Figure 2). This structure pattern is very different from the solvation shells formed by monoatomic inorganic cations, such as Li[+]

(Figure 4). The two clearly defined maxima with zero density between them suggest two well-ordered solvation shells.

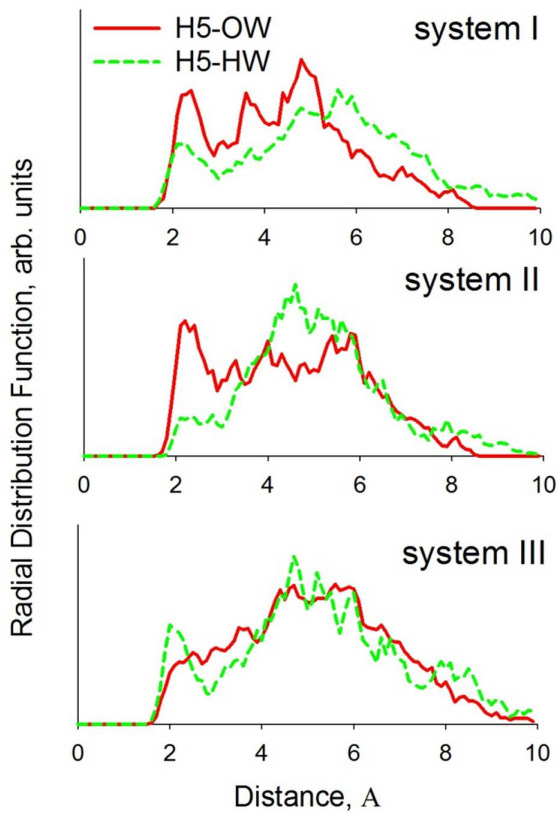

Figure 3. Radial distribution functions computed for the intrinsically acidic hydrogen atom of the imidazole ring and oxygen atoms of water (OW) and of methanol (OH) in systems I, II, and III.

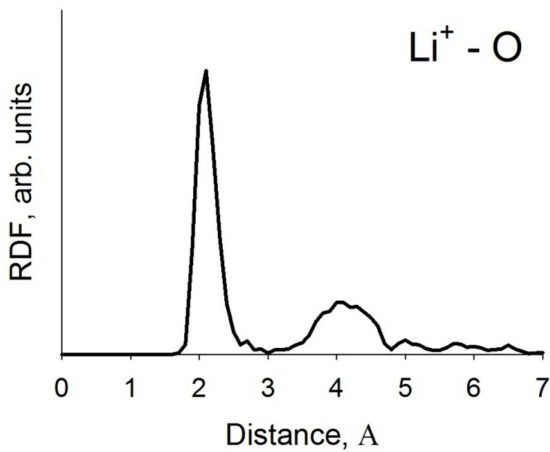

Figure 4. Radial distribution function (RDF) in system IV derived for the nucleus of the lithium atom (ion) and oxygen atoms of the methanol molecules forming the first and second solvation shells. The first peak of the RDF is located at 2.1 Å and clearly indicates a strong hydrogen bond. The second solvation shell is located at 4.0 Å.

The first solvation shell of the imidazolium cation is predominantly populated by water molecules, whereas most methanol molecules get shifted to a poorly defined second solvation shell. Since both polar solvents exhibit similar hydrogen bonding potential, the current solvation observation may get explanation based on the structure of hydrogen bonds in the vicinity of the solvated cation. Figure 5 summarizes possible hydrogen bonds involving oxygen and hydrogen atoms of water and methanol molecules. These RDFs do not account for the hydrogen bond between $MMIM^+$ and solvent molecules, which was considered in detail above. Remarkably, methanol—methanol hydrogen bonding is significantly weaker, as compared to water-water one and even to water-methanol one. This conclusion does not depend on the composition of the solvent mixture (systems I-III). Water engenders a strong hydrogen bond with a length of 1.7 Å in all three solutions/mixtures, whereas a large fraction of the OH-HO distances in methanol is more than 4 Å. The OH-HW is located at 1.7 Å, 2.2 Å and in between, indicating that water is able to insert between the two methanol molecules and maintain a hydrogen-bonded network. That is, water separated methanol pair, in analogy to solvent separated ion pair, may be observed. The analysis of distances favors such a supposition and provides certain explanation why water dominates in the first solvation shell of $MMIM^+$.

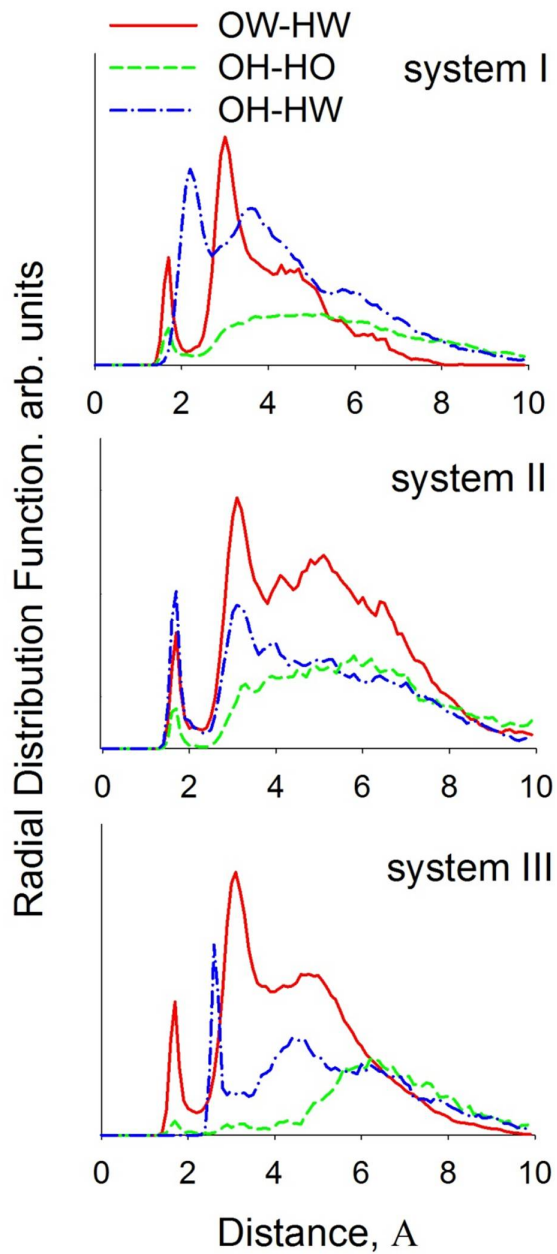

Figure 5. Radial distribution functions computed for atom pairs, which create hydrogen bonds: OW-HW (water – water hydrogen bond); OH-HO (methanol – methanol hydrogen bond); OH-HW (water – methanol hydrogen bond).

The total energy value must be conserved during an entire molecular dynamics simulation. If the integration time-step is set too large, this may be a permanent channel for energy leakage. Such energy leakage is dangerous, since it may provide implicit simulation artifacts, which are often very difficult to locate and, hence, remove. Figure 6 provides a necessary benchmark for an optimal integration time-step within the velocity Verlet algorithm. The time-steps of 0.5, 1.0, and 2.0 fs were probed. It was found that only 0.5 and 1.0 fs offer acceptable energy conservation, while the time-step of 2.0 fs is too large for accurate results. The energy dissipation likely occurs due to fast oscillations of the hydrogen–oxygen covalent bonds both in water and methanol molecules. Setting larger atomic masses of the respective hydrogen atoms may partially solve the problem, but will simultaneously change the corresponding momentum of inertia and rotational dynamics.

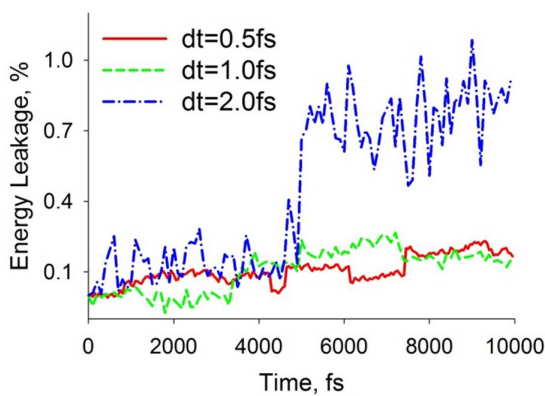

Figure 6. Total energy conservation in system II while increasing an integration time-step. The energy leakage is provided as a percentage of total potential energy of this system at the beginning of each simulation. The simulations have been carried out in the constant energy ensemble (no temperature coupling). Prior to constant-energy simulations, the corresponding system was equilibrated at 333 K with a time-step of 0.5 fs. Whereas 0.5 and 1.0 fs offer an acceptable energy conservation level, the time-step of 2.0 fs demonstrates poorer accuracy. Therefore, the production simulations (Figures 1-5) apply a time-step of 1.0 fs.

**Conclusions**

This work reports semiempirical molecular dynamics simulations of the 1,3-dimethylimidazolium cation solvated by water and methanol. Three compositions of the water-methanol mixture were considered. Structures of polar molecular solvents in the vicinity of the cation were characterized in terms of pair radial distribution function. The intrinsically acidic hydrogen atom of the imidazole ring is responsible for strong ion-molecular couplings. Water exhibits preferential solvation of $MMIM^+$, but does not exclude methanol completely from the first solvation shell. The performed analysis suggests that successful solvation ability of water deals with its ability to penetrate into the hydrogen-bonded network created by methanol molecules. Both solvents are successful in engendering a moderately strong hydrogen bond (2.1 Å) with 1,3-dimethylimidazolium.

Normalization of the reported RDFs deserves a few explicit comments. The simulated systems are non-periodic due to the technical limitation imposed by PM7. In particular, the Slater basis set is not efficient to work with periodic boundary conditions. Additionally, periodicity in the case of relatively small systems is difficult to justify, since periodic images cannot substitute a long-range order of the liquid phase, which is beyond the scope of my current research interest. The simulated systems do not have boundaries and, hence, they do not have volumes. The computed RDFs can be normalized with respect to the number of MD time-steps, but they cannot be, strictly speaking, normalized with respect to each species' density. This poses a problem for comparison of the RDFs with one another. Constant volume approximation can be accepted for the sake of comparison within the same MD system. Hence, we formally define the volume as unity and compute a relative density of each interaction center based on the number of these interaction centers in the system. Thus, the density of H5 equals to 1, the density of H4 equals to 2, the density of OW in system I equals to 9 and so on. Now the available information permits to compare RDFs to one another. Recall that comparison is valid only within the same system and the same MD simulation. The height of RDFs may make physical sense as long as the integral of the RDF provides a running coordination number of certain particle with respect to another particle. Otherwise, only positions of maxima and shape of the peaks provide physically meaningful information, which can be altogether used to unveil three-dimensional model of ionic solvation shell or other structural pattern.

The results reported here assist in fostering progress in chemistry and chemical engineering thanks to better understanding of ion-molecular interactions and solvation behavior in the case of organic cation. Water and methanol exhibit similar internal structure and distribution of electron density within a single molecule. It was impossible to hypothesize, based on the available data, which molecule prevails in the first solvation shell of MMIM$^+$. A natural supposition would be that composition of the first coordination sphere of MMIM$^+$ is determined solely by concentration of water or methanol in the solvating mixture. This ad hoc hypothesis was confirmed only partially. Indeed, the composition of the first solvation shell depends on the composition of the mixture. However, water molecules exhibit somewhat better ability to solvate MMIM$^+$ due to joining hydrogen bonded network of the neighboring methanol molecules.

The investigation additionally illustrates power, robustness, and reliability of PM7-MD for the consideration of multi-component chemical systems, while usage of empirical pairwise potentials is associated with uncertainty due to the neglected quantum electronic effects.

**Acknowledgments**

This investigation has been partially supported by a research grant from CAPES (Coordenação de Aperfeiçoamento de Pessoal de Nível Superior, Brasil) under "Science Without Borders" program. Profs. Eudes Eterno Fileti and Thaciana Malaspina are acknowledged for care, hospitality, and remote library access. Brazil is acknowledged for sunny winter months and relaxed world view.